\documentclass[aps,prb,twocolumn,showpacs,superscriptaddress]{revtex4-1}

\usepackage{graphicx} 

\begin{document}

\title
{Frictional figures of merit for single layered nanostructures}

\author{S. Cahangirov}
\affiliation{UNAM-National Nanotechnology Research Center, Bilkent University, 06800 Ankara, Turkey}
\affiliation{Institute of Materials Science and Nanotechnology, Bilkent University, Ankara 06800, Turkey}
\author{C. Ataca}
\affiliation{UNAM-National Nanotechnology Research Center, Bilkent University, 06800 Ankara, Turkey}
\affiliation{Institute of Materials Science and Nanotechnology, Bilkent University, Ankara 06800, Turkey}
\affiliation{Department of Physics, Bilkent University, Ankara 06800, Turkey}
\author{M. Topsakal}
\affiliation{UNAM-National Nanotechnology Research Center, Bilkent University, 06800 Ankara, Turkey}
\affiliation{Institute of Materials Science and Nanotechnology, Bilkent University, Ankara 06800, Turkey}
\author{H. Sahin}
\affiliation{UNAM-National Nanotechnology Research Center, Bilkent University, 06800 Ankara, Turkey}
\affiliation{Institute of Materials Science and Nanotechnology, Bilkent University, Ankara 06800, Turkey}
\author{S. Ciraci}\email{ciraci@fen.bilkent.edu.tr}
\affiliation{UNAM-National Nanotechnology Research Center, Bilkent University, 06800 Ankara, Turkey}
\affiliation{Institute of Materials Science and Nanotechnology, Bilkent University, Ankara 06800, Turkey}
\affiliation{Department of Physics, Bilkent University, Ankara 06800, Turkey}

\date{\today}

\begin{abstract}
We determined frictional figures of merit for a pair of layered honeycomb nanostructures, such as graphane, fluorographene, MoS$_2$ and WO$_2$ moving over each other, by carrying out ab-initio calculations of interlayer interaction under constant loading force. Using Prandtl-Tomlinson model we derived critical stiffness required to avoid stick-slip behavior. We showed that these layered structures have low critical stiffness even under high loading forces due to their charged surfaces repelling each other. The intrinsic stiffness of these materials exceed critical stiffness and thereby avoid the stick-slip regime and attain nearly dissipationless continuous sliding. Remarkably, tungsten dioxide displays much better performance relative to others and heralds a potential superlubricant. The absence of mechanical instabilities leading to conservative lateral forces is also confirmed directly by the simulations of sliding layers.
\end{abstract}

\pacs{68.35.Af, 62.20.Qp, 81.40.Pq} \maketitle

Advances in atomic scale friction\cite{atomic,persson,renaissance} have provided insight on dissipation mechanisms. The stick-slip phenomena is the major process, which contributes to the dissipation of the mechanical energy through sudden or non-adiabatic transitions between bi-stable states of the sliding surfaces.\cite{prandtl,tomlinson,tomanek,buldum1} During a sudden transition from one state to another, the velocities of the surface atoms exceed the center of mass velocity sometimes by orders of magnitudes.\cite{review} Local vibrations are created thereof evolve into the non-equilibrium system phonons via anharmonic couplings\cite{gurevich} within picoseconds.\cite{buldum2} In specific cases, even a second state in stick-slip can coexist.\cite{buldum1}

\begin{figure}
\includegraphics[width=8.5cm]{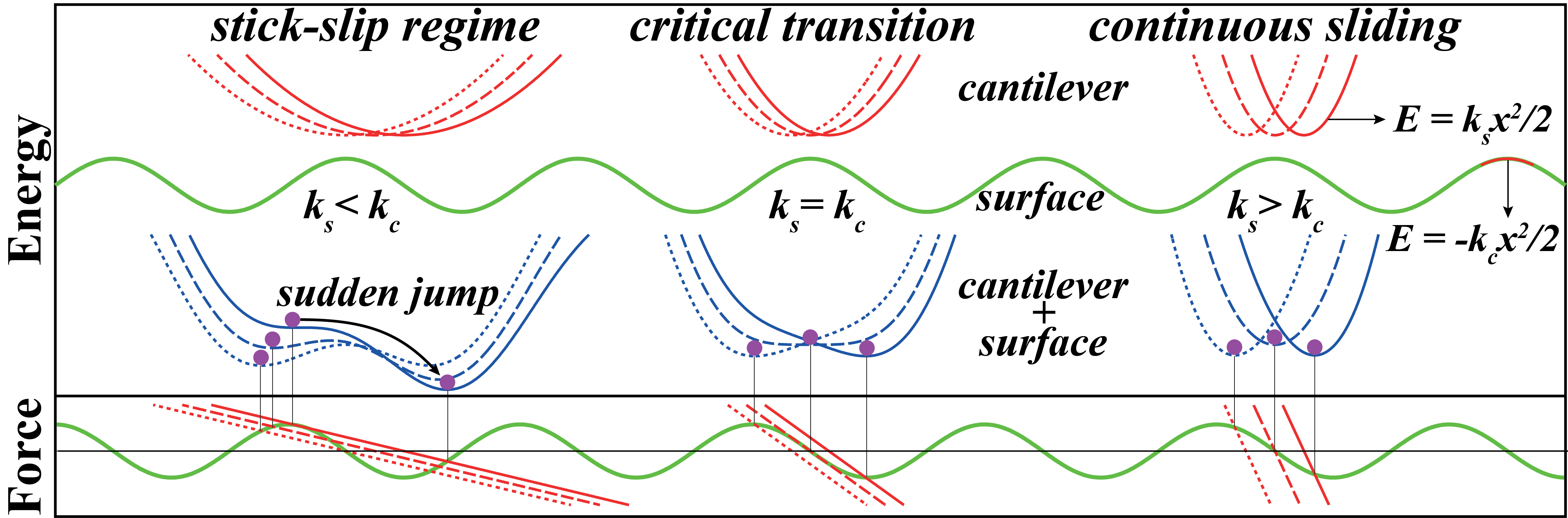}
\caption{(Color Online) Schematic representation of stick-slip regime (left), critical transition (middle) and continuous sliding regime (right) in Prandtl-Tomlinson model. Upper part: the potential energy curves of the surface (green lines) and of the tip(+cantilever) (red lines) ; lower part: force variation of the surface (green lines) and of the tip (red lines). Blue lines represent the potential energy of the tip and surface. The magenta dot shows the position of the tip on the surface, while its other end is positioned at the minimum of the parabola shown with red lines in the upper part. The dotted, dashed and solid lines correspond to three different tip positions moving to the right.}
\label{fig1}
\end{figure}

In Fig.\ref{fig1}, two regimes of sliding friction are summarized within the framework of Prandtl-Tomlinson model,\cite{prandtl,tomlinson,review} where an elastic tip(+cantilever) moves over a sinusoidal surface potential. The curvature of this potential at its maximum gives the value of the critical stiffness $k_c$. If the intrinsic stiffness of the tip $k_s$ is higher than this critical stiffness \textit{i.e.} $k_{s}/k_{c}>1$, the total energy of the tip-surface system always has one minimum. The sliding tip gradually follows this minimum, which results in the continuous sliding regime. Conversely, if the tip is softer than the critical value, then it is suddenly slipped from one of the bi-stable states to the other. This slip event can be activated by thermal fluctuations even before the local minimum point becomes unstable.\cite{gnecco} Experimentally, using friction force microscope, Socoliuc \textit{et al.}\cite{meyer} showed that the transition from stick-slip regime to continuous sliding attaining ultralow friction coefficient can be achieved by tuning the loading force on the contact.

In this Letter, the sliding friction between two same pristine layers of nanostructures, such as graphane,\cite{graphane,hasan1} fluorographene,\cite{fluorographene,hasan2} molybdenum disulfide\cite{can1} and tungsten dioxide,\cite{can2} (abbreviated according to their stoichiometry as CH, CF, MoS$_2$ and WO$_2$ respectively) is investigated using the Density Functional Theory\cite{dft}. We find that these nanostructures avoid stick-slip even under high loadings and execute continuous sliding. Consequently, the sliding occurs without friction that would originate from the generation of non-equilibrium phonons. Our approach mimics the realistic situation, where the total energy and forces are calculated from first-principles as two-dimensional (2D) two layers undergo a 3D sliding motion under a constant (normal) loading force. This is the most critical and difficult aspect of our study. In this respect, our results provide a 3D rigorous \textit{quantum mechanical} treatment for the 1D and empirical Prandtl-Tomlinson model.\cite{prandtl,tomlinson}

\begin{figure}
\includegraphics[width=8.5cm]{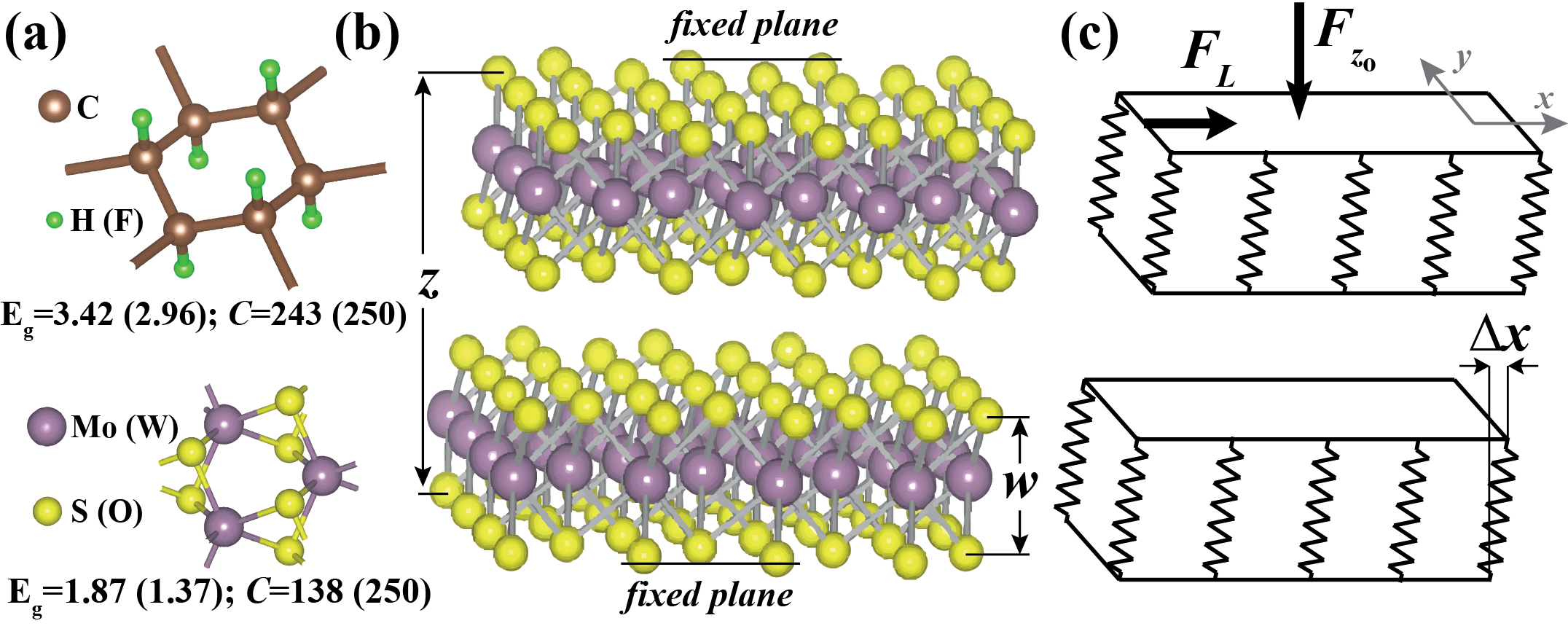}
\caption{(Color Online) (a) Ball and stick model showing the honeycomb structure of graphane CH (fluorographene CF) (top) and MoS$_2$ (WO$_2$) (bottom). Calculated values of energy gaps $E_g$ and in-plane stiffness $C$ are also given in units of eV and $J/m^2$ respectively. (b) Two MoS$_2$ layers sliding over each other have the distance $z$ between their outermost atomic planes. (c) Each layer is treated as a separate elastic block. Lateral $F_L$ and normal (loading) $F_{z_{o}}$ forces, the shear of bottom atomic plane relative to top atomic plane in each layer $\Delta x(y)$, and the width of the layer $w$, are indicated.}
\label{fig2}
\end{figure}

The nanostructures considered in the present study are recently discovered insulators having honeycomb structure, which can form suspended single layers as well as  multilayers. The unusual electronic, magnetic and elastic properties of these layers have been the subject of recent numerous studies. In particular, they have large band gaps to hinder the dissipation of energy through electronic excitation and have high in-plane stiffness\cite{apl,hasan2,can1,can2} ($C=(1/A)\partial^{2}E_{s}/\partial \epsilon^{2}$, i.e. the second derivative of the strain energy relative to strain per unit area, $A$ being the area of the unit cell). Analysis based on the optimized structure, phonon and finite temperature molecular dynamics calculations demonstrate that each suspended layer of these nanostructures are planarly stable.\cite{hasan1,hasan2,can1,can2} In graphane, positively charged three hydrogen atoms from the top side and another three from the bottom are bound to the alternating and buckled carbon atoms at the corners of hexagons in graphene to form a uniform hydrogen coverage at both sides (See Fig.\ref{fig2}(a)). Recently synthesized CF\cite{fluorographene} is similar to CH, but F atoms are negatively charged. Tribological properties of carbon based fluorinated structures have been the focus of interest.\cite{miyake,dubois} In the layers of MoS$_2$ or WO$_2$, the plane of positively charged transition metal atoms is sandwiched between two negatively charged outer S or O atomic planes.  It was shown that MoS$_2$ structure can have ultralow friction.\cite{martin} Theoretically, the static energy surfaces are calculated during sliding at MoS$_2$(001) surfaces.\cite{liang} Apparently, the interaction energy between two single layers of these nanostructures is mainly repulsive due to charged outermost planes except very weak Van der Waals attractive interaction around the equilibrium distance. In Fig.\ref{fig2}, each layer being a large 2D sheet consisting of three atomic planes mimics one of two sliding surfaces. In practice, sliding surfaces can be coated by these single layer nanostructures as one achieved experimentally.\cite{acsnano}

We consider two layers of the same nanostructures in relative motion, where the spacing $z$ between the bottom atomic plane of the bottom layer and the top atomic plane of the top layer is fixed. Here the frictional behavior of the system is dictated mainly by C-H(F), Mo-S and W-O bonds and their mutual interactions. These layers are represented by periodically repeating rectangular unit cells. We calculate the value of the equilibrium lattice constants, which increase as $z$ decreases. For each value of $z$ the fixed atomic layer at the top is displaced by $x$ and $y$ on a mesh within the quarter of the rectangular unitcell. Then all possible  relative positions (displacements) between fixed atomic layers are deduced using symmetry. At each mesh point all atoms of the system except those of fixed top and bottom planes are relaxed and the total energy of the system $E_T(x,y,z)$ (comprising both layers) is calculated. We have also derived $\Delta x(x,y,z)$ and $\Delta y(x,y,z)$ data which correspond to the shear (deflection) from the equilibrium position of the relaxed atomic planes relative to the fixed atomic plane of the same layer as illustrated in Fig.\ref{fig2}(c). The matrices of these data are arranged for each nanostructure using the mesh spacing of $\sim$ 0.2~$\AA$ in $x$ and $y$ directions. The forces exerting on the displacing top layer in the course of relative motion of layers are calculated from the gradient of the total energy of the interacting system, namely $\vec{F}(x,y,z)=-\vec{\nabla} E_T(x,y,z)$ at each mesh point ($x,y$). These forces are in agreement with the resultant of the atomic forces calculated for the top layer using Hellman-Feynman theorem. Eventually, the matrices of all data, namely $E_T(x,y,z)$, $\Delta x(x,y,z)$, $\Delta y(x,y,z)$ and $\vec{F}(x,y,z)$ are made finer down to mesh spacing of $\sim$~0.05~$\AA$ using spline interpolation.

The properties affecting the friction between layers should be derived under a given constant loading force. First of all we preset the value of applied loading, $F_{z_{o}}$, which corresponds to the operation pressure when divided by the cell area $A$, namely $\sigma_N=F_{z_{o}}/A$. We obtain the normal force from $F_z(x,y,z)=-\partial{E_T(x,y,z)}/\partial{z}$  and for each $x$ and $y$ we calculate the value of $z$ where $F_z(x,y,z)=F_{z_{o}}$ and abbreviate it as $z_{o}(x,y)$. Then by using spline interpolation in $z$ direction we calculate  the $x$ and $y$ dependence of $F_{x_{o}}[x,y,z_{o}(x,y)]$ and $F_{y_{o}}[x,y,z_{o}(x,y)]$, as well as $\Delta x_o[x,y,z_o(x,y)]$ and $\Delta y_o[x,y,z_o(x,y)]$ for a  given $F_{z_{o}}$. The lateral force is then $\vec{F}_{L}[x,y,z_{o}(x,y)]=F_{x_{o}}\hat{i}+F_{y_{o}}\hat{j}$. Integrating the lateral force over the rectangular unitcell we obtain,

\begin{eqnarray} \nonumber
E_{I}[x,y,z_o(x,y)]=\int_0^x \int_0^y  \vec{F}_{L}(x,y,z_o(x,y))  \cdot \vec{\mathrm{d}r}
\end{eqnarray}

where $E_{I}[x,y,z_o(x,y)]$ is the interaction energy for displacement $(x,y)$ in the cell under applied constant loading force $F_{z_{o}}$. It should be noted that $E_{I}$ is different from $E_{T}(x,y,z)$ (but $E_{I} \rightarrow E_T$ for $z>>1$) and is essential to reveal the friction coefficient. Contour plots of $E_I$ of two sliding MoS$_2$ layers calculated for $\sigma_N$=15 GPa are shown in Fig.\ref{fig3}(a) and those of CH, CF, WO$_2$ in Supplemental-Material-A.\cite{supp-a} The profile of $E_{I}$ is composed of hills arranged in a triangular lattice. These hills correspond to the relative positions when the charged atoms of adjacent layers have the minimum distance. The hills are surrounded by two kind of wells. The difference between these two wells is enhanced with increasing pressure. The wells form a honeycomb structure and are connected to each other through the saddle points (SP). When the layers are moved over each other they will avoid the relative positions corresponding to the hills. For example, if the layers are pulled in the $y$-direction they will follow the curved $F_x=0$ path passing through the wells and SP but not the straight one passing through the hills as shown in the Fig.\ref{fig3}(b). This makes SP very important because moving from one well to the adjacent one requires to overcome the barriers at these points. We note that the critical stiffness can be calculated from the curvature of $E^{o}_{I}$, which is obtained by subtracting the strain energies of two sliding MoS$_2$ layers, namely $E^{o}_{I}= E_{I}-k_{s}(\Delta x_{o}^{2}+\Delta y_{o}^{2})$ and by replacing $x$ by $x-2\Delta x_{o}$. While the SP serves as a barrier in the direction joining the nearby wells it acts as a well in the perpendicular direction joining the hills. Since we are interested in the curvature of the SP in the former direction we have made a plot along the $F_y=0$ line which passes through the hill, the wells and the SP in between as shown in the Fig.\ref{fig3}(b). We derive two critical stiffness values from $E^{o}_{I}$ curve for a given normal loading force; namely $k_{c1}$ at the SP and $k_{c2}$ at the hill by fitting the curve at the maxima of the barriers to a parabola. Although the hills will be avoided during sliding motion the curvature at these points are calculated for completeness. In Fig.\ref{fig3} (c) the variation of $k_{c1}$ and $k_{c2}$ of CH, CF, MoS$_2$ and WO$_2$ with loading pressure $\sigma_N$ is presented. Generally, the critical stiffness, in particular $k_{c1}$ is low due to repulsive interaction between sliding layers. This facilitates the transition to continuous sliding.

Next we calculate the intrinsic stiffness $k_s$ of individual MoS$_2$ layers using the force and the displacement data. For each $x$ and $y$ the lateral forces $F_{x_o}[x,y,z_o(x,y)]$ and $F_{y_o}[x,y,z_o(x,y)] $ versus the displacements $\Delta x_o[x,y,z_o(x,y)]$ and $\Delta y_o[x,y,z_o(x,y)]$, respectively are plotted. As shown in Fig.\ref{fig3}(b) as inset, this data falls on a straight line having a negative slope as expected from Hook's law of elasticity. We note that the elastic properties of layers having honeycomb structure is uniform and is independent of the direction of displacement and force.\cite{apl} The magnitude of the slope, $k_{s} = -F_{x(y)_{o}}/\Delta x(y)_o$ gives us the stiffness of the layers.\cite{unit} Calculated intrinsic stiffness values of CH, CF, MoS$_2$ and WO$_2$ in the range of $\sigma_N$ from 5GPa to 30 GPa are found to be 6.15$\pm$0.15 eV/\AA$^{2}$, 4.5 eV/\AA$^{2}$, 10.0$\pm$0.3 eV/\AA$^{2}$~ and 15.2$\pm$0.3 eV/\AA$^{2}$, respectively. Clearly, these values of $k_s$, in particular those of MoS$_2$ and WO$_2$ are rather high.

\begin{figure}
\includegraphics[width=8.5cm]{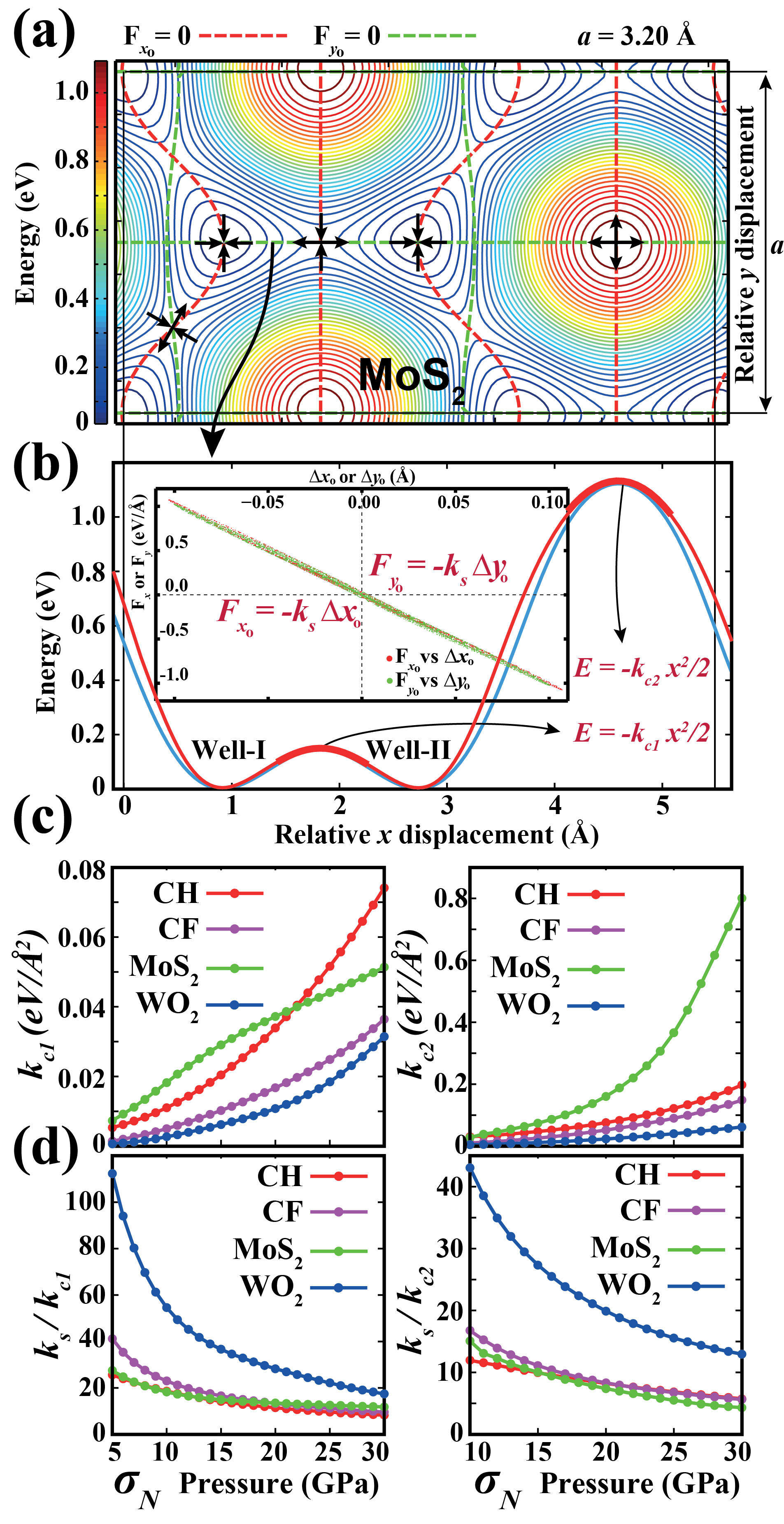}
\caption{(Color Online)  (a) The contour plot of interaction energy $E_{I}$ of two sliding layers of MoS$_2$. The zero of energy is set to $E_{I}[0,0,z_o(0,0)]$. The energy profile is periodic and here we present the rectangular unitcell of it. The width of this unitcell in $y$-direction is equal to the lattice constant $a$ of the hexagonal lattice. Forces in $x$- ($y$-) direction is zero along the red (green) dashed lines, respectively. There are several points at which the lateral force $\vec{F_L}$, is zero. The arrows at these critical points indicate the directions where the energy decreases. (b) The energy profiles of $E_I$ (blue line) and $E^{o}_I$ (red line) along the horizontal line with $F_y=0$ for MoS$_2$. The inset presents force versus shear values along $x$- and $y$-directions for each mesh point by red and green dots, respectively, which fall on the same line. Loading pressure in all cases is $\sigma_N$=15 GPa. (c) The variation of $k_{c1}$ and $k_{c2}$ with loading pressure. (d) The variation of the ratios of $k_s/k_{c1}$ and $k_s/k_{c2}$, i.e. frictional figures of merit with loading pressure calculated for CH, CF, MoS$_2$ and WO$_2$.}
\label{fig3}
\end{figure}

Based on the discussion at the beginning, the ratios $k_s/k_{c1}$ and $k_s/k_{c2}$ give us a dimensionless measure of performance of our layered structures in sliding friction. When these ratios are above two (since both layers in relative motion contribute), the stick-slip process is replaced by continuous sliding, whereby the dissipation of mechanical energy through phonons is ended. Under these circumstances the friction coefficient diminish, if other mechanisms of energy dissipation were neglected. For this reason one may call these ratios as a frictional figures of merit of the layered materials. In Fig.\ref{fig3} (d) we present the variations of the ratios $k_s/k_{c1}$ and $k_s/k_{c2}$ with normal loading forces. Even for very large $\sigma_N$, $k_s/k_{c1} >2 $ and $k_s/k_{c2} >2$. For usual loading pressures, the stiffness of MoS$_2$, CF and CH is an order of magnitude higher than corresponding critical values. Interestingly, for WO$_2$ this ratio can reach to two orders of magnitudes at low pressures. The absence of mechanical instabilities has been also tested by performing extensive simulations of the sliding motion of layers in very small displacements. C-H, C-F, Mo-S and W-O bonds in each case of two layers in relative motion under significant loading force did not display the stick-slip motion.

\begin{figure}
\includegraphics[width=8.5cm]{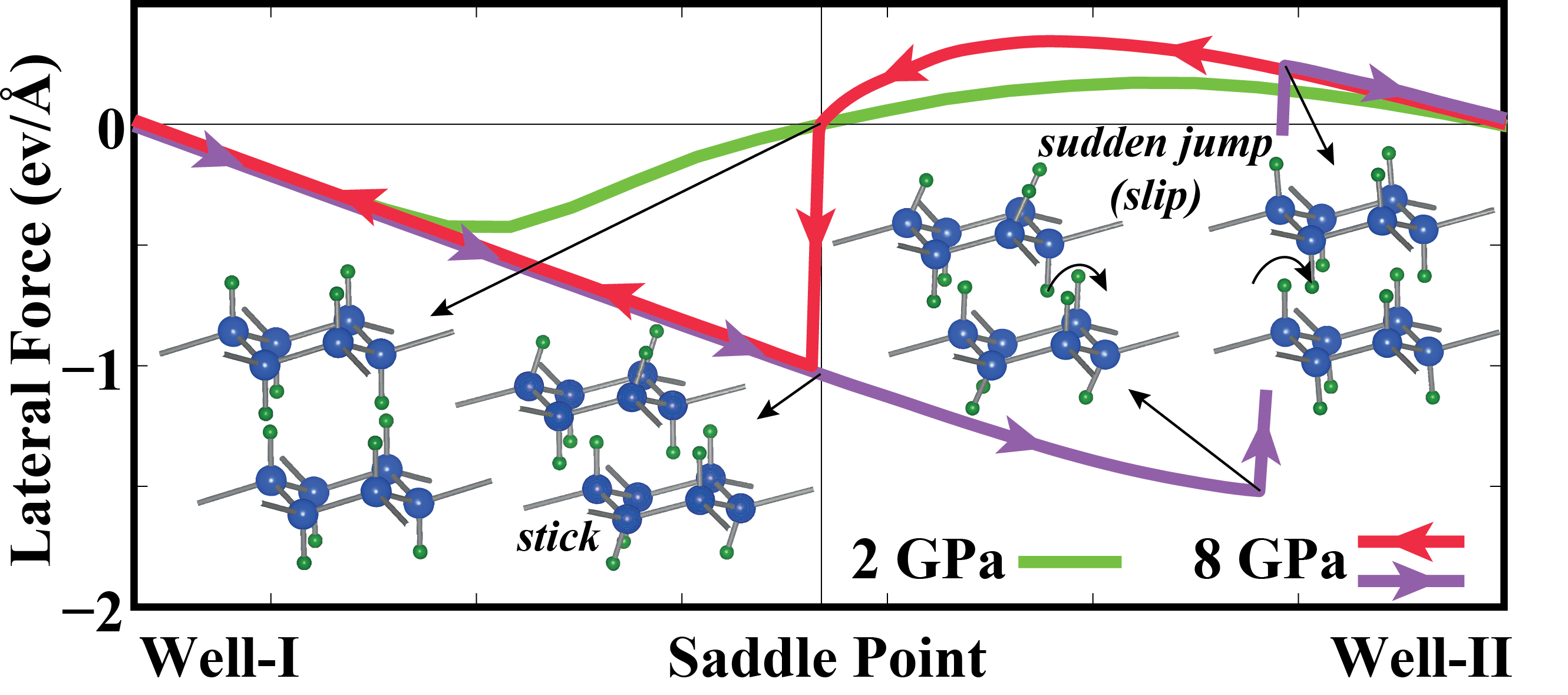}
\caption{(Color Online) Calculated lateral force variation of two single layer SiH under two different $\sigma_N$. The top layer is moving to the right or to the left between two wells. Atomic positions of two SiH layers in stick and slip stages are shown by inset. The movement of SiH layers under loading pressure of $\sigma_N$ = 8~GPa is presented in Supplemental-Material-B.\cite{supp-b}}
\label{fig4}
\end{figure}

Conversely, we now examine the sliding of two silicane\cite{silicane} layers (abbreviated as SiH and composed of silicene\cite{seymur} saturated by hydrogen atoms from both sides, like graphane) with $k_s=2.1 \pm 0.1$~eV/$\AA^2$ for 2~GPa $\le \sigma_N \le$ 8~GPa. This is an interesting material because the onset of stick-slip occurs already at low loading pressures and exhibits a pronounced asymmetry in the direction of sliding between two wells. In Fig.\ref{fig4} we present the lateral force variation calculated for two different loading pressures. For small loading pressure, $\sigma_N$=2~GPa the stick-slip is absent since approaching the SP from Well-I, the curvature is $k_{c,I}=0.28$~eV/$\AA^2$ and from Well-II it is $k_{c,II}=0.16$ eV/$\AA^2$, thus $k_s/k_{c,I or II} > 2$ for both directions. Whereas, once the pressure is raised to $\sigma_N=$8~GPa stick-slip already governs the sliding friction, since $k_{c,I}$ reaches 1.38~eV/$\AA^2$. Interestingly, since $k_{c,II}$ is only 0.28~eV/$\AA^2$ for $\sigma_N=$8~GPa, going from Well-II to Well-I a slip event occurs at SP. Eventually, one sees in Fig.\ref{fig4} a hysteresis in the variation of $F_L$ leading to energy dissipation.

Earlier, the sliding motion of the diamond like carbon coatings exposed to hydrogen plasma resulted in a very low friction coefficient.\cite{erdemir} Ultralow friction was attributed to repulsive Coulomb forces between DLC films facing each other in sliding. However, when exposed to open air in ambient conditions, positively charged H atoms was replaced by negatively charged O and hence the uniformity in the charging was destroyed. In the present study, graphane coating is reminiscent of the hydrogenated DLC and accordingly is found to have ultralow friction, but vulnerable to degradation by oxygen atoms. Unlike graphane coating, WO$_2$ coating consists of negatively charged oxygens and hence immune to oxidation.

In conclusion, using a criterion for the transition from stick-slip to dissipationless continuous sliding regime, which is calculated from the first-principles, we showed that two sliding layered nanostructures, such as CH, CF, MoS$_2$ and WO$_2$, execute continuous sliding with ultralow friction. The minute variation of the amplitude of the interaction potential due to the repulsive interaction, as well as stiff C-H(F), Mo-S and W-O bonds underlie the frictionless sliding predicted in the present study. Our predictions put forward an important field of application as ultralow friction coating for the layered honeycomb structures, which can be achieved easily to hinder energy dissipation and wear in sliding friction.

This work is supported by TUBITAK through Grant No:108T234 also by EFS EUROCORE programme FANAS. All the computational resources have been provided by TUBITAK ULAKBIM, High Performance and Grid Computing Center (TR-Grid e-Infrastructure).


\begin{thebibliography}{15}
\bibitem{unit}
Note that, normally the stiffness is defined as stress over strain and has units of energy per volume. Here we only need the ratio of material stiffness to the critical stiffness and should have the same units. The critical stiffness was calculated as second order spatial derivative of energy in the unitcell and it has units of energy per unitcell per unit area. As defined above, the stiffness of layers, $k_s$, also has units of energy per unitcell per unit area.
\end{thebibliography}
\end{document}